\documentclass{IAU-TrA}
\usepackage{graphicx}

\def\kms{\ifmmode{\rm km\thinspace s^{-1}}\else km\thinspace s$^{-1}$\fi}
\def\ms{\ifmmode{\rm m\thinspace s^{-1}}\else m\thinspace s$^{-1}$\fi}
\def\cms{\ifmmode{\rm cm\thinspace s^{-1}}\else cm\thinspace s$^{-1}$\fi}

\title[RADIAL VELOCITIES]     
{}

\author[DIVISION~IX / COMMISSION 30]   
{}

\pubyear{2012}
\volume{Volume XXVIIIA}
\pagerange{001--008}
\jname{Transactions IAU, Volume XXVIIIA}
\editors{Ian Corbett, ed.}
\begin{document}

\maketitle

{\bf

\large
\begin{tabbing}
\hspace*{65mm}       \=                                              \kill
COMMISSION~30        \> RADIAL VELOCITIES                     \\
                     \> {\small\it VITESSES RADIALES}         \\
\end{tabbing}

\normalsize

\begin{tabbing}
\hspace*{65mm}       \=                                              \kill
PRESIDENT            \> Guillermo Torres                      \\
VICE-PRESIDENT       \> Dimitri Pourbaix                      \\
PAST PRESIDENT       \> Stephane Udry                         \\
ORGANIZING COMMITTEE \> Geoffrey W.\ Marcy,                   \\ 
                     \> Robert D.\ Mathieu, Tsevi Mazeh,      \\
                     \> Dante Minniti, Claire Moutou,         \\
                     \> Francesco Pepe, Catherine Turon,      \\   
                     \> Tomaz Zwitter                         \\
\end{tabbing}

\noindent
COMMISSION~30 WORKING GROUPS
\smallskip

\begin{tabbing}
\hspace*{65mm}      \=                                               \kill
Div.~IX / Commission~30 WG \> Stellar Radial Velocity Bibliography   \\
Div.~IX / Commission~30 WG \> Radial Velocity Standards              \\
Div.~IX / Commission~30 WG \> Catalogue of Orbital Elements of       \\
                           \> Spectroscopic Binary Systems (SB9)     \\
\end{tabbing}

\bigskip

\noindent
TRIENNIAL REPORT 2009-2012
}

\smallskip

\firstsection 

\section{Introduction}

The past three-year period has seen steady efforts to collect large
numbers of radial-velocity (RV) measurements, as well as important
applications of radial velocities to astrophysics.  Improvements in
precision continue to be driven largely by exoplanet research. A
workshop entitled ``Astronomy of Exoplanets with Precise Radial
Velocities'' took place in August of 2010 at Penn State University
(USA), and was attended by some 100 researchers from around the
world. The meeting included thorough discussions of the current
capabilities and future potential of the radial velocity technique, as
well as data analysis algorithms to improve precision at visible and
near-infrared wavelengths.

\smallskip

While most of these discussions were focused on the search for and
characterization of exoplanets, it is clear that more classical
applications of radial velocities are also benefiting from the
improvements, as evidenced by recent work on binary stars described
herein. Below is a summary of other activity in the field of radial
velocities during this triennium. Due to space limitations, we include
only a selection of efforts and results in this area.

\section{Large-scale radial-velocity surveys {\rm (T.\ Zwitter and G.\
Torres)}}

The RAdial Velocity Experiment (RAVE; {\tt
http://www.rave-survey.org}) is an ongoing international collaboration
of $\sim$60 scientists from nine countries, led by M.\ Steinmetz from
the AIP in Potsdam. It is continuing to use the UK Schmidt telescope
at the Australian Astronomical Observatory to record a large unbiased
sample of stellar spectra selected only by their $I$-band
magnitude. During this triennium RAVE publicly released its full pilot
survey (\cite{Siebert:11a}), which contains 86,223 RV measurements for
81,206 stars in the southern hemisphere. In addition, stellar
parameters for 42,867 of the stars were published. Altogether RAVE has
already collected over 500,000 spectra, with approximately 10\% of the
observing time being devoted to repeat observations. The mean radial
velocity error is $\sim$2~\kms, and 95\%\ of the measurements have an
internal error better than 5~\kms. This can be combined with distances
based on 2MASS photometry, spectroscopically determined values of the
stellar parameters, and stellar isochrone fitting (see
\cite{Breddels:10, Zwitter:10, Burnett:11}).  Such distances are
accurate to $\sim$20\% and cluster around 300~pc for dwarfs and 1 or
2~kpc for giants.

\smallskip

This collection of information allows comprehensive studies of the
kinematics of our part of the Galaxy, as well as its structure and
formation history. RAVE is suited to searching for stellar streams,
some of which are remnants of dwarf galaxies that merged with the
Milky Way during galaxy formation. A new stream, dubbed the Aquarius
stream, is an example of such remnants that can be found with RAVE
(\cite{Williams:11}). Another kind of stellar streams, known as moving
groups, are born inside our Galaxy.  New members of nearby moving
groups have been found in RAVE (\cite{Kiss:11}), and the survey
promises to reveal more in the future. RAVE allows for efficient
searches for the very first stars (\cite{Fulbright:10}), and enables
the detection of interesting trends in the motions of the stars in the
vicinity of the Sun (\cite{Siebert:11b}). The survey is well suited to
study our Galaxy's thick disk. Two recent studies from RAVE
(\cite{Wilson:11, Ruchti:11}) have focused on uncovering its origin.
The survey will continue in 2012.

\smallskip

Another large-scale survey released during this triennium that
includes radial-velocity measurements (albeit of low precision) for
vast numbers of stars is SEGUE (Sloan Extension for Galactic
Understanding and Exploration). A paper by \cite{Yanny:09} describes
these spectroscopic results, which are based on some 240,000
low-resolution ($R \sim 1800$) spectra of fainter Milky Way stars down
to a magnitude limit of $g \approx 20.3$. One of the goals is to
enable studies of the kinematics and populations of our Galaxy and its
halo. The RV precision varies from 4~\kms\ at the bright end ($g
\approx 18$) to 15~\kms\ at the faint end. In addition to the
velocities, atmospheric parameters including effective temperature,
surface gravity, and metallicity were derived for the stars with
suitable signal-to-noise ratios. The individual spectra along with
associated parameters are publicly available as part of the Sloan
Digital Sky Survey Data Release~7.

\section{The role of radial-velocity measurements in studies of
stellar angular momentum evolution and stellar age {\rm (S.\ Meibom)}}

Radial-velocity measurements with multi-object spectrographs have
played a critical role in defining the mile-posts that are the
foundation for much of our understanding of the time-evolution of
stars. These mile-posts are star clusters --- coeval, cospatial, and
chemically homogeneous populations of stars over a range of masses for
which the age can be determined well by fitting model isochrones to
single cluster members in the color-magnitude diagram (CMD). However,
the inherent qualities of clusters can only be fully exploited if
pure samples of kinematic members are identified and characterized.
This can be accomplished most securely and effectively with
radial-velocity measurements (e.g., \cite{Geller:08, Hole:09}).

\smallskip

Recent dedicated photometric surveys for stellar rotation periods in
young clusters have begun to see dependencies of stellar rotation on
stellar age and mass. These dependencies guide our understanding of
the angular momentum evolution of FGK dwarfs by determining the mass-
and time-dependence of their rotation periods. Over the past three
years such surveys have been combined with radial-velocity surveys for
cluster membership and binarity in open clusters with different ages,
revealing well-defined relations between stellar rotation period,
color (mass), and age not previously discernible (\cite{Meibom:09a,
Meibom:11a, Meibom:11b}). These relations offer crucial new
constraints on internal and external angular momentum transport and on
the evolution of stellar dynamos in late-type stars of different
masses.

\smallskip

Furthermore, stellar rotation has emerged as a promising and
distance-independent indicator of age (``gyrochronology'';
\cite{Kawaler:89, Barnes:03, Barnes:07}), and open clusters fulfill an
important role in calibrating the relation between age, rotation, and
mass.  Indeed, open clusters can define a surface in the
three-dimensional space of stellar rotation period, mass, and age,
from which the latter can be determined from measurements of the
former two. It is critical, however, to establish the cluster ages
from CMDs in which non-members have been removed and single members
identified. It is also important to identify short-period binaries
where tidal mechanisms may have modified the stellar rotation. Radial
velocities are an efficient and proven technique to identify both
single and short-period binary members. The tight mass-rotation
relations seen in clusters over the past three years reflect the
powerful combination of time-series spectroscopy for cluster
membership and time-series photometry for rotation periods.

\section{Radial velocities in open clusters {\rm (R.\ Mathieu)}}

Studies of kinematic membership and binarity in open clusters based on
radial-velocity measurements have a long history. During this
triennium the WIYN Open Cluster Study (WOCS; \cite{Mathieu:00}) has
continued to acquire intermediate-precision ($\sigma_{\rm RV} =
0.4$~\kms) radial-velocity measurements on its core open clusters.
Currently the project has in hand a total of more than 60,000
measurements of some 11,800 stars in the open clusters M34, M35, M37,
M67, NGC\,188, NGC\,2506, NGC\,6633, NGC\,6819, and NGC\,7789. Some of
these data and associated results have already appeared in the
literature (\cite{Geller:08, Geller:09, Geller:10, Hole:09,
Meibom:09a, Meibom:09b, Meibom:11b}).

\smallskip

Particularly notable progress has been made on understanding the
nature of blue stragglers in the open cluster NGC\,188
(\cite{Mathieu:09, Geller:11}).  Sixteen of the 21 blue stragglers are
spectroscopic binaries.  These binaries have a remarkable eccentricity
versus log period distribution, with all but two having periods within
a decade of 1000 days.  The two short-period binaries are
double-lined, one of which comprises \emph{two} blue stragglers.  A
statistical analysis of the single-lined binary mass functions shows
the secondary mass distribution to be narrowly confined around a mass
of 0.5~$M_{\odot}$.  The combination of these results strongly
suggests a mass-transfer origin for the blue stragglers, leaving behind
white dwarf companions.  However, the shortest period binaries are
certainly the product of dynamical encounters, leaving open the
possibility of collisional origins for those blue stragglers.

\section{Toward higher radial-velocity precision {\rm (F.\ Pepe,
C.\ Moutou, C.\ Lovis)}}

Broadly speaking, this period has been characterized by three general
trends regarding precise radial-velocity measurements as applied to
exoplanet research. Firstly, the precision has been pushed to its
limits such that very small RV signals even below 1~\ms\ have now been
detected. This capability has revealed a large population of
super-Earths and Neptunes, demonstrating that they are common around
solar-type stars in the Milky Way (\cite{Howard:11, Mayor:11}).
Secondly, Doppler shift measurements have become a tool complementary
to other techniques, and in particular to the transit method of
detecting exoplanets. And thirdly, RVs are moving into the
near-infrared domain. The combination of red wavelengths and very high
spectral resolution has only become available in recent years, but has
already brought on previously unavailable opportunities for the
observation of stars that are very young, very active, or of very late
spectral type, and opened up possibilities for the detection of
planetary signatures among those stars.

\smallskip

Recent developments have demonstrated that at the few \ms\ level the
star is not necessarily the limiting factor, and that there is good
reason to aim for \emph{sub}-\ms\ instrumental precision (see, e.g.,
\cite{Pepe:11}) provided the star is chromospherically quiet and that
photon noise is not the limit. Considerable progress has been made on
instrumental issues. One of the limiting factors has been the
non-uniform illumination of the spectrograph, where even the use of
(circular) fibers does not remove this problem entirely.  Non-circular
fibers have shown great promise for their scrambling properties,
although much of this work has not yet appeared in the literature. One
exception is the study by \cite{Perruchot:11} with octagonal-section
fibers. Using these devices it has been possible to improve the RV
precision on the SOPHIE spectrograph mounted on the 1.93\,m OHP
telescope from about 8~\ms\ to 1.5~\ms. The other important factor
limiting instrumental precision is the wavelength calibration. The two
main techniques used for this (thorium-argon lamps, and the iodine
cell method) have a limited wavelength coverage, suffer from line
blending, and have other drawbacks (including large dynamic range for
the lines and limited lifetime of hollow-cathode lamps, and light
absorption as well as sensitivity to ambient conditions for the iodine
cell). The use of laser frequency combs as a path to achieving \cms\
precision has been explored for several years, and a number of these
systems are now under development for both visible (\cite{Osterman:07,
Steinmetz:08, Li:09}) and infrared wavelengths (\cite{Osterman:11,
Schettino:11}). Challenges still remain, but the expectation is that
these devices will be available on several telescopes around the world
on a timescale of a few more years.

\smallskip

The problems posed by the spectrograph illumination and wavelength
calibration will likely be solved soon. Present-generation
spectrographs are already implementing solutions to those challenges
based on the technologies mentioned above. At the \cms\ level,
however, stellar ``jitter'' will still be an important source of
error.  Current efforts to overcome this have focused on filtering the
stellar noise contribution ($p$ modes, granulation, activity) by
applying optimal observation strategies (see, e.g.,
\cite{Dumesque:11}). Future planet search programs requiring extremely
high precision will likely have to pre-select targets with very low or
very well-known stellar jitter, so that these effects either have
minimal impact on the RVs, or can be modeled and removed. And of
course, beating down photon noise in the search for Earth-like planets
will require ever larger telescopes, or restricting the searches to
relatively bright stars.

\smallskip

Achieving very high velocity precision in the near-infrared has so far
lagged behind the optical regime. Performance at the \ms\ has not yet
been achieved, although 5~\ms\ has been demonstrated in a few cases
(e.g., \cite{Bean:10, Figueira:10}). The problems to be overcome
include the treatment of telluric lines, detector technology, and
cryogenic optics.

\smallskip

Several new radial-velocity instruments are presently under
construction that should come online in the next few years. A
non-exhaustive list with an indication of the wavelength regime,
telescope on which they will be mounted, and expected first-light date
includes HARPS-N (visible, TNG, 2012), PEPSI (visible, LBT, 2012),
GIANO (IR, TNG, 2012), HZPF (IR, HET, 2013), CARMENES (visible-IR,
3.6\,m Calar Alto, 2014), SPIROU (IR, CFHT, 2015), and ESPRESSO
(visible, VLT, 2016).

\section{High-precision radial velocities applied to studies of binary
stars {\rm (G.\ Torres)}}

As indicated above, one of the procedures used in exoplanet research
for ensuring high precision in the radial-velocity measurements relies
on an iodine cell in front of the spectrograph slit to track
instrumental drifts and changes in the point-spread function that
normally lead to systematic errors (see, e.g., \cite{Marcy:92,
Butler:96}). Some years ago \cite{Konacki:05} extended the iodine
technique to composite spectra, showing that precisions of a few tens
of \ms\ can be reached in selected double-lined spectroscopic
binaries. This enables considerably higher precision to be obtained
for the masses of binary stars than has usually been achieved (see
also earlier work by \cite{Lacy:92}).

\smallskip

A recent study by \cite{Konacki:10} focused on a handful of favorable
(nearly edge-on) binaries, and combined spectroscopy with
long-baseline interferometric observations, which yield the
inclination angle of the orbit, to achieve record precision for one of
their systems, HD\,210027. Relative errors in the masses are as low as
0.066\%, the smallest obtained for any normal star. Other studies by
the same group have also reached very small uncertainties
(\cite{Helminiak:11a, Helminiak:11b}), made possible by the much
improved velocities using their technique. The precision of the masses
of HD\,210027 rivals that of the best known determinations in double
neutron star systems, measured by radio pulsar timing.

\section{Doppler boosting effect {\rm (T.\ Mazeh)}}

In the last two years a new type of stellar radial-velocity
measurement has emerged, based on the photometric beaming (aka Doppler
boosting) effect. This causes the bolometric flux of a star to
increase or decrease as it moves toward or away from the observer,
respectively. The magnitude of the beaming effect is approximately $4
V_r/c$, where $V_r$ is the stellar radial velocity and $c$ is the
speed of light, and is therefore on the order of $10^{-3}$ to
$10^{-4}$ of the stellar intensity for a solar-type star with a
stellar secondary and a period of 10 days or so.

\smallskip

While the beaming effect had been observed previously from the ground
in one or two very favorable cases (e.g., \cite{Maxted:00}), the
availability of a quarter of a million very precise, continuous light
curves produced by the \emph{CoRoT} and \emph{Kepler} missions has
opened the door to the detection of new binary systems by this method
(see also \cite{Loeb:03, Zucker:07, Faigler:11a}).  Seven new
\emph{non-eclipsing} binaries with orbital periods between 2 and 6
days have already discovered by this effect in the \emph{Kepler} data,
and were confirmed by classical spectroscopic radial-velocity
measurements (\cite{Faigler:11b}). The effect has now also been seen
in ground-based photometry of two extremely short period double white
dwarf eclipsing binaries with periods of 5.6 and 0.2 hours
(\cite{Shporer:10, Brown:11}), as well as in other eclipsing systems
observed by \emph{Kepler} that also contain white dwarfs
(\cite{vanKerkwijk:10, Carter:11, Breton:11}).

\section{Working Groups {\rm (H.\ Levato, G.\ Marcy, D.\ Pourbaix)}}

Below are the reports of the three active working groups of Commission
30. Their efforts are focused on providing a service to the
astronomical community at large through the compilation of a variety
of information related to radial velocities.

\subsection{WG on Stellar Radial Velocity Bibliography (Chair:
H.\ Levato)}

This WG is a very small one that was created with the purpose of
continuing the cataloging of the bibliography of radial velocities of
stars made by Mme Barbier in successive catalogues until her
retirement in 1990 (see \cite{Barbier:90}).

\smallskip

The new compilation was started late in 1990. The first version of the
catalogue after the retirement of Mme Barbier was published for the
1991--1994 triennium.  The catalogue is updated every six months at
the following web page:

\smallskip
\noindent {\tt http://www.icate-conicet.gob.ar/basededatos.html}
\smallskip

During the 2009--2011 period the WG searched 33 journals for papers
containing measurements of the radial velocities of stars. As of
December 2010 a total of 198,063 entries had been cataloged. By the
end of 2011 this is expected to increase to about 285,000 records.  It
is worth mentioning that at the end of 1996 the number of entries was
23,358, so that in 15 years the catalogue has grown by more than an
order of magnitude.  The main body of the catalogue includes
information about the technical characteristics of the instrumentation
used for the radial velocity measurements, and comments about the
nature of the objects.

\smallskip
 
The future of radial velocities is becoming very attractive and the
same time more complex. Large numbers of new radial velocity
measurements are expected to be published, and it may be necessary to
discuss if the present approach is the best way to keeping a record of
the bibliography of radial velocity measurements.

\subsection{WG on Radial Velocity Standards (Chair: S.\ Udry)}

During this triennium significant progress has been made towards
establishing lists of stars that can serve as radial-velocity
standards, to a much higher level of precision than lists that have
been used in the past. Two main efforts have taken place.

\smallskip

One was summarized by \cite{Crifo:10}, who report the compilation of
an all-sky list of 1420 relatively bright (mostly $V \approx 6$--10)
stars developed specifically for use by the Gaia project, but which is
of course very useful to the broader community. The list is based
largely on measurements published by \cite{Nidever:02},
\cite{Nordstrom:04}, and \cite{Famaey:05}.  The radial velocities of
most these stars are believed to be accurate at the $\sim$300~\ms\
level, and a large fraction of them are being re-observed at higher
precision with modern instruments (SOPHIE, NARVAL, CORALIE; see
\cite{Chemin:11}). It is expected that the accuracy will be improved
to 100 or possibly 50~\ms\ when this task is concluded. A link to this
list of potential standards is available on the Commission web page.

\smallskip

A parallel effort reported by \cite{Chubak:11} has been carried out by
the California Planet Search group using the HIRES spectrometer on the
Keck~I telescope. They present radial velocities with an accuracy (RMS
compared to present IAU standards) of 100~\ms\ for 2086 stars of
spectral type F, G, K, and M based on some 29,000 spectra.  Additional
velocities are presented for 132 RV standard stars, all of which
exhibit constant radial velocity for at least 10 years, with an RMS
less than 10~m\,s$^{-1}$. All velocities were measured relative to the
solar system barycenter and are placed on the velocity zero-point
scale of \cite{Nidever:02}. They contain no corrections for convective
blueshift or gravitational redshift.  An innovation was to determine a
secure wavelength zero-point for each spectrum by following the
suggestion of Roger Griffin in using telluric lines (the origin of the
iodine cell concept).  Specifically, they used the telluric A and B
bands at 7594--7621\,\AA\ and 6867--6884\,\AA, respectively, which
were present in all of the spectra. This allows to correct for small
changes in the CCD position, the spectrometer optics, and guiding
errors for the specific observation of the program star.

\smallskip

There is a significant overlap between the lists of \cite{Crifo:10}
and \cite{Chubak:11}, providing excellent radial velocity integrity
for the stars in common. It is expected that the combination of these
lists will serve as standards for studies of long-period binary stars,
star cluster dynamics, and for surveys of the chemical and dynamical
structure of the Galaxy such as SDSS, RAVE, Gaia, APOGEE, SkyMapper,
HERMES, and LSST.

\subsection{WG on the Catalogue of Orbital Elements of Spectroscopic
Binaries (SB9) (Chair: D.\ Pourbaix)}

At the 2000 General Assembly in Manchester, a WG was set up to work on
the implementation of the 9th Catalogue of Orbits of Spectroscopic
Binaries (SB9), superseding the 8th release of \cite{Batten:89} (SB8).
SB9 exists in electronic format only.  The web site ({\tt
http://sb9.astro.ulb.ac.be}) was officially released during the summer
of 2001.  This site is directly accessible from the Commission 26 web
site, from BDB (in Besan\c{c}on), and from the CDS, among others.

\smallskip

Substantial progress have been made since the last report, in
particular in the way complex multiple systems can be uploaded
together with their radial velocities.  The way data weights can be
supplied has also been improved.

\smallskip

As of this writing the SB9 contains 3039 systems (SB8 had 1469) and
3784 orbits (SB8 had 1469).  A total of 623 papers were added since
August 2000, with most of them coming from \emph{outside} the WG. A
significant number of papers with orbits still await uploading into
the catalogue.  According to the ADS, the release paper
(\cite{Pourbaix:04}) has received 152 citations since 2005.  This is
about three times more than the old Batten et al.\ catalogue over the
same period, with the SB8 still being cited in the current literature.

\smallskip

The important work of cross-checking the identification of systems is
carried out by the CDS (Strasbourg).  Indeed, with the SBC9 identifier
now added to SIMBAD, each new release of the SB9 tar ball is cross
checked for typos prior to integration at the CDS.  Whereas some of
these mistakes are ours, some authors share the responsibility as
well.  Users have also helped in pinning down some problems.

\smallskip

Although this work is very welcome by the community (about 500--1000
successful queries received every month, with 50 distinct IP addresses
over the past month) and some tools have been designed to make the job
of entering new orbits easier (input file checker, plot generator,
etc.), the WG still suffers from a serious lack of manpower.  Few
colleagues outside the WG spontaneously send their orbits (though they
are usually happy to send their data when we asked).  Any help from
authors, journal editors, etc., is therefore very welcome.  Uploading
an orbit into SB9 also means checking it against typographical errors.
In this way we have found a number of mistakes in published solutions.
Sending orbits to SB9 prior to publication (e.g., at the proof stage)
would therefore be a way to prevent some mistakes from making their
way into the literature.

\vspace{3mm}
 
{\hfill Guillermo Torres}

{\hfill {\it president of the Commission}}


\end{document}